# Analytical modeling and Dynamics of Multi-Domains in Negative-Capacitance MFIS-FETs.

Nilesh Pandey and Yogesh Singh Chauhan
Department of Electrical Engineering, Indian Institute of Technology Kanpur, Kanpur, INDIA
E-mail: pandeyn@iitk.ac.in; chauhan@iitk.ac.in

*Abstract—* Analytical modeling and dynamics of multidomain in metal-ferroelectric-insulator-semiconductor (MFIS)-FETs are presented in this paper. The formation of multi-domain (MD) leads to oscillations in the conduction band in the channel and periodicity in the local electric field in the ferroelectric region. The impact of 2-D local electric field on the MD switching is captured in the model using the domain wall velocity concept. The optimum values of oxide thickness, ferroelectric thickness and channel length are calculated which corresponds to mono-domain device operation. Deviation from the optimum device parameters causes the transition of mono-domain state to multi-domain state in the ferroelectric. This work can be used as a guideline for designing MFIS-NCFETs, which provides the device parameters that leads to monodomain state in the MFIS-NCFET.

## I. INTRODUCTION AND METHOLOGY

The formation of multi-domains (MD) in ferroelectric (FE) is an inevitable process during the minimization of system energy [1]. To simplify the analysis, most of the of MFIS-NCFET studies consider only mono-domain condition in the ferroelectric region [2]-[7]. Very few studies have included the impact of MD impact on the ferroelectric. However, these reported papers have neglected the study of state transition, domain wall velocity and the optimization of device. Besides, these studies are based on the numerical solutions of phase field equations [8]-[12].

The divergence in the spontaneous polarization is responsible for the formation of MD in the FE region [1]. To study the impact of MD on FE, $180^0$ strip domain pattern is assumed in the ferroelectric region, which is a valid and realistic assumption, for the temperature less than the Curie temperature [1]. Using this assumption, the polarization associated with each domain can be expressed in the form of a Fourier series (1). The 2-D Poisson equation with the divergence of spontaneous polarization term is solved using Green's function approach [15]. This solution is completely analytical and explicit in nature. The obtained potential function of the FE region ($\phi_{MD}(x,y)$) is used to calculate the domain wall velocity ($V_{dw}$) and the switching of domains is captured by $V_{dw}$ concept [13]-[14]. The domain period ($T_D$) is not known in $\phi_{MD}(x,y)$. Hence, to calculate the $T_D$, first, we consider the equilibrium condition $(V_{gs} = V_{ds} = 0\ V)$ in the MFIS-NCFET. Later, the net thermodynamic energy of the system is minimized to calculate the equilibrium domain period ($T_{D0}$) [16]. The obtained $T_{D0}$ is used in (4) to derive the complete expression of $\phi_{MD}(x,y)$ which valid only for the equilibrium state. The next step is to obtain the value of the domain period and domain wall velocity for a nonzero gate/drain voltage. An iterative procedure is used to calculate the alignment of each domain with the 2-D local field (see **Fig. 1**). The obtained angle is further used to calculate the voltage-dependent polarization and the domain period for the nonzero applied gate/drain voltage. Finally, the $\phi_{MD}(x,y)$ is updated with the new domain period (for nonzero gate/drain voltage), and, $V_{dw}$ is calculated for any value of applied $V_{gs}$ and $V_{ds}$.

## II. RESULTS AND DISCUSSION

Fig. 2 shows the equilibrium domain formation in MFIS-NCFET with the default parameters used in this paper. Fig. 3(a), shows the conduction band profiles plotted at the mid-channel, for various domain periods ($T_D$). As the number of domains increases, the oscillations in $E_c$ increase, due to increment in the periodicity of adjacent positive and negative charges, in the FE region. Fig. 4(a), shows the drain current for the various $T_D$ values, indicates that smaller number of domains (larger $T_D$) reduces the OFF current. Fig. 4(b), shows the subthreshold slope (SS) as a function of $T_D$, for various ferroelectric thickness ($t_{fe}$) values, as the number of domains increases, SS starts to degrade, due to MD effect. Maximum two domains exist for $T_D > $ L/2, hence, SS begins to saturate at its minimum value, as $T_D$ approaches to the L/2 limit.

Fig. 5(a), shows the switching of domain's polarization with variation in the applied gate voltage $(V_{gs})$. As $V_{gs}$ increases, the vertical direction ferroelectric electric field increases, and this in turn, increases the domain wall velocity in the lateral direction [13]. Moment of the upward-facing domains in the x-direction causes gradual reduction in the number of downward-facing domains and this is defined as the switching of domains [14]. Due to this switching of domains oscillations in $E_c$, decrease as $V_{gs}$ increases in Fig. 5(b). Additionally, the period of upward-facing domains ($x_2$) also increases with the increment in $V_{gs}$, and this is another way to define the domain switching in the FE region. Fig. 6, shows the variation in lateral direction E-field of FE $(E_{x,fe})$ with $V_{gs}$. Similar to Fig. 5(b) higher $V_{gs}$ increases $x_2$. Hence, at higher $V_{gs}$ smaller number of oscillations in the $E_{x,fe}$ is observed.

Fig. 7(a), shows the variation of lateral-directional polarization $(P_{szx})$ of the domains with drain voltage $(V_{ds})$. Domains near drain end experience strong lateral E field, hence, their orientation angel $(\theta_z)$ is larger than to domains near source end (see Fig. 1). Therefore, the magnitude of $P_{szx}$ increases gradually, as domain approaches near drain end. On the other hand, as can be seen in Fig. 7(b), the magnitude of vertical-direction polarization $(P_{szy})$ decreases near drain end, due to reduction in the y-direction polarization vector component. Further, $V_{ds}$ also contributes to domain wall velocity and, hence, the switching of domains also takes place with variation in $V_{ds}$, as evident from Fig. 7(a) and (b). Fig. 8 and Fig. 9 show the variation $E_c$ and $E_{x,fe}$ with $V_{ds}$. As $V_{ds}$ increases, $T_D$ increases due to the gradual alignment of the polarization in the lateral direction (switching of domains) or along lateral-E field.

According to [13], domain wall velocity increase exponentially with increases in the net 2-D local electric field. Hence, switching of the MD increases, as net E field increases. Fig. 10(a) and Fig. 10(b) show the $V_{dw}$ for various $L$ and $V_{ds}$ values. Shorter $L$ or higher $V_{ds}$ enhances the lateral E field which increases the magnitude of $V_{dw}$. However, as $t_{ox}$ increases $V_{dw}$ decreases, due to reduced vertical E field at higher insulator thickness. Therefore, the switching of domains can be significantly controlled by the 2-

D local E field. Also, each domain in the FE region has different $V_{dw}$, due to variation in the local E field (see Fig. 6 and Fig. 9).

According to [16] at the higher $t_{ox}$ monodomain state shows an abrupt transition to multidomain state. The maximum value of $t_{ox}$, at which the transition takes place is calculated using (9) by putting $T_D = L$, which represents the extremum of the thermodynamic potential ($\xi$) [17]. Subsequently, $\xi$ is minimized with $t_{ox}$ ($\partial \xi / \partial t_{ox} = 0$), to find out the maximum value of $t_{ox}$, at which the system can exists in the monodomain state, with the minimum thermodynamic energy. Fig. 11(a), shows that as channel length increases (for fixed $t_{ox}$), monodomain states starts to shift towards the multidomain state. Therefore, a long channel device is more likely to have multidomain states and, hence, reduced NC effect. Fig. 11(b), shows the transition of states with variation in $t_{fe}$. At larger $t_{fe}$, multidomain state transitions into monodomain state, which is the opposite compared with L. Fig. 11(a) and (b), can be used as device design guidelines, to obtain the optimum values of $t_{fe}, t_{ox}$ and L to operate devices in the monodomain region (stronger NC region).

## III. CONCLUSION

Formation of multi-domains is an essential process to minimize the total system energy. The switching of multi-domains does not directly depend on the gate/drain voltage. Instead, 2-D local electric field is responsible for the switching. Each domain in the FE region moves with a different $V_{dw}$ due to different E field in each domain. Each domain has dissimilar polarization vector which depends on the local E field of the domain. The transition from monodomain to multidomain depends mainly on $t_{fe}, t_{ox}$ and the L. The optimum set of physical parameters, to operate the device in monodomain state can be predicted by the model.

**Spontaneous Polarization Profile ($180^0$ strip domains with the domain wall thickness $d_w$)**

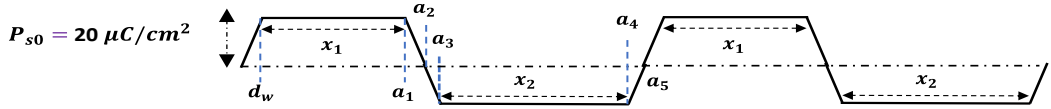

$$P_s(x) = P_{s0} \sum_j \frac{(a_j + c_j)cos(k_j x) + (b_j + d_j)sin(k_j x)}{k_j^2} \quad (1)$$

$$a_j = cos(k_j d_w) - 1 + cos(k_j a_1) - cos(k_j a_2); c_j = cos(k_j a_2) - cos(k_j a_3) + cos(k_j a_2) - cos(k_j a_3) + cos(k_j a_5) - cos(k_j a_4)$$
$$b_j = sin(k_j d_w) - sin(k_j a_2) + sin(k_j a_1); d_j = sin(k_j a_2) - sin(k_j a_3) + sin(k_j a_5) - sin(k_j a_4) \quad (2)$$

**2-D Potential modeling**

$$\frac{\partial \phi^2}{\partial x^2} + \frac{\partial \phi^2}{\partial y^2} = \frac{\Delta . P_s}{\epsilon_{fe}} \quad (3)$$

**Ferroelectric Region**

$$\phi_{MD}(x,y) = \frac{2}{t_{fe}} \sum_{n=1}^{\geq 5} \frac{sin(k_n^I y)}{sinh(k_n^I L)} \{A_1 sinh(k_n^I (L-x)) + A_2 sinh(k_n^I (L-x))\} + \phi_{pol}(x,y) + \frac{2}{L} \sum_{m=1}^{\geq 10} \frac{sin(k_m x)sinh(k_m y)(P_s^m - D_1^m)}{k_m cosh(k_m t_{fe}) \epsilon_{fe}}$$

$$\frac{2}{L} \sum_{m=1}^{\geq 10} \frac{sin(k_m x)cosh(k_m (t_{fe} - y))(V_{gs} - \phi_m + \phi_s)(1+(-1)^{m+1})}{k_m cosh(k_m t_{fe})} + \frac{2}{L} \sum_{m=1}^{\geq 10} \frac{sin(k_m x)sinh(k_m y)(P_s^m - D_1^m)}{k_m cosh(k_m t_{fe}) \epsilon_{fe}} \quad (4)$$

$$\phi_{pol}(x,y) = \frac{4 P_s}{d_w T_D \epsilon_{fe} t_{fe}} \sum_n^{\geq 5} \frac{sin(k_n^I y)}{k_n^{I^2} sinh(k_n^I L)} \sum_j \frac{(b_j + d_j)I_1 - (a_j + c_j)I_2}{k_j k_n^{I^2} + k_j^3} \quad (5)$$

$$I_2 = k_n^I sin(k_j x)sinh(k_n^I L) - k_n^I sinh(k_n^I x)sin(k_j L); I_1 = k_n^I cos(k_j x)sinh(k_n^I L) - k_n^I sinh(k_n^I (L-x)) - k_j cos(k_j L)sinh(k_n^I x) \quad (6)$$

**Oxide region**

$$\phi_{ox}(x,y) = \frac{2}{t_{ox}} \sum_{n=1}^{\geq 5} \frac{cos\left(k_n^{II}(t_{fe} + t_{ox} - y)\right)}{sinh(k_n^{II} L)} \{B_1 sinh(k_n^{II}(L-x)) + B_2 sinh(k_n^{II}(L-x))\}$$

$$+ \frac{2}{L} \sum_{m=1}^{\geq 10} \frac{sin(k_m x)}{k_m sinh(k_m t_{ox}) \epsilon_{ox}} \left(D_1^m cosh\left(k_m(t_{fe} + t_{ox} - y)\right) - D_2^m cosh\left(k_m(t_{fe} - y)\right)\right) \quad (7)$$

**Channel region:**

$$\varphi_{si}(x,y) = \frac{E_g}{2q} + \frac{x}{L} V_{ds} + \frac{2}{L} \sum_{m=1}^{\geq 10} \frac{sin(k_m x) D_2^m \left(cosh\left(k_m(t_{fe} + t_{ox} + t_{si} - y)\right) + cosh\left(k_m(t_{fe} + t_{ox} - y)\right)\right)}{k_m sinh(k_m t_{si}) \epsilon_{si}} \quad (8)$$

Where, $A_1, A_2, B_1, B_2$ are the boundary gap that are evaluated by considering linear potential profiles along the x=0, and x=L line. $D_1^m$ and $D_2^m$ are Fourier series coefficients which are evaluated by the potential continuity conditions at $(x, t_{fe})$ and $(x, t_{fe}+t_{ox})$ boundary interfaces and $P_s^m = \int_0^L P_s(x) sin(k_m x) dx$.

## Modeling of domain swithchig and domain wall velocity

Initially, the system is considered is to be in the equilibrium state. Hence the work produces by the voltage sources and the energy of the electric field is expressed as ($\xi$) [16]:

$$\xi = E_{LGD} - 2q\phi_{fe}(x_1, 0) + \int E_{fe}^2/8\pi \quad (9)$$

Where, $E_{LGD}$ is standard Landau-Ginzburg-Devonshire functional at zero electric field and $E_{fe}$ is the 2-D electric field in ferroelectric.

$$\frac{\partial \xi}{\partial x_1} = 0, (a) \rightarrow T_{D0} \sim 2x_1 (x_1 = x_2), \rightarrow P_{sz}^{i-1}(x) \quad (10)$$

$V_{gs} = V_{ds} = 0$ is the equilibrium condition of domain formations, due to the gradient in spontaneous polarization. Eq. (9) is the condition of minimization of net thermodynamic potential [17], which is used to calculate the equilibrium domain period ($T_{D0}$).

$$\phi_{fe}(x, y) \rightarrow E_{fe}(x_d, t_{fe}/2), (11) \rightarrow \Delta = (x_1 - x_2) \rightarrow$$

$$\frac{d\Delta}{dt} = -2V_{dw} \sim -2\exp\left(-\frac{E_{act}}{E_{net}^{i-1}}\right) [13], \rightarrow T_D^{i-1} \rightarrow$$

$$\theta_z^{i-1} = \tan^{-1}\left|\left(\frac{E_{x,fe}^{i-1}}{E_{y,fe}^{i-1}}\right)\right| \quad (12)$$

Where, t is the switching time of the domain [14], $E_{x,fe}$ and $E_{y,fe}$ are ferroelectric electric fields in $x, y$ directions respectively. $x_d = zx_1/2 + (z-1)x_2/2$, z is an integer that represents the $z^{th}$ domain in the ferroelectric region and, $z_{max} = 2L/T_{D0}$. $V_{dw}$, is the domain wall velocity [13], which depends on the 2-D local electric fields. The switching of domains is captured by the moment of domain wall in the FE region.

The value of $\theta_z$ is obtained from (12) used to determine the x and y direction polarization components as:

$$P_{yz}^i(x, \theta) = P_{sz}^{i-1} \sum_z \cos(\theta_z^{i-1}); P_{xz}^i(x, \theta) = P_{sz}^{i-1} \sum_z \cos(\theta_z^{i-1}). \quad (13)$$

$$P_{sz}^i = \sqrt{(P_{yz}^i)^2 + (P_{xz}^i)^2} \quad (14)$$

Eq. (14) is used in (10) to calculate the updated $P_{sz}^i$. Subsequently, (12) is updated to obtain new $\theta_z$. An iterative procedure among (10), (12), and (14) is used to obtain the accurate orientation and magnitude of the $z^{th}$ domain polarization. Note that, due to hyperbolic functions in the electric fields, only 3-4 iterations needed for the convergence.

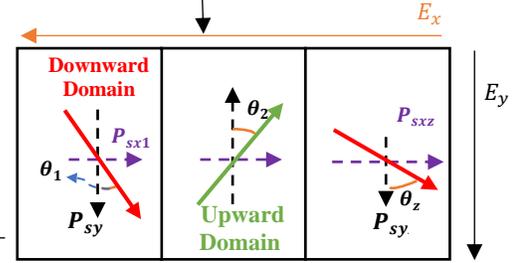

**Fig. 1:** Polarization vector orientation angel ($\theta_z$) for the applied gate and drain voltages. The angle $\theta_z$, increases towards the drain side due to increment in x directional polarization component.

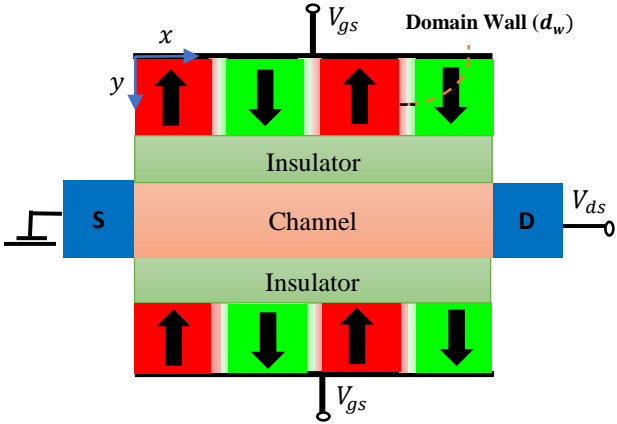

**Fig. 2:** Equilibrium domain formation in MFIS-NCFET. Default Parameters: $t_{ox}$= 0.5 nm, $t_{fe}$= 6 nm, $\phi_m$ = 4.4 eV, S/D doping =$1 \times 10^{20} cm^{-3}$, $V_{gs}$= 0 V and $V_{ds}$= 0.7 V. The material parameters of HfO$_2$ ferroelectric are taken from [14].

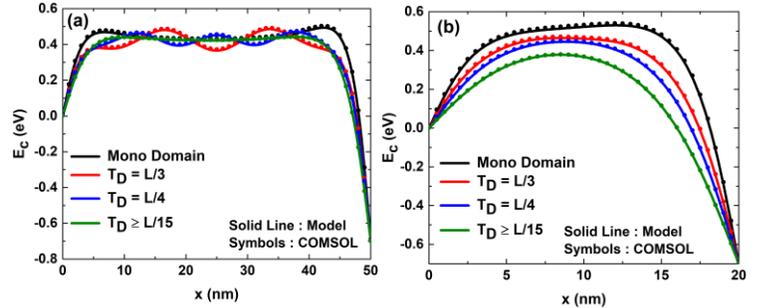

**Fig. 3:** Conduction band energy, plotted in the middle of the channel for various domain periods ($T_D$). (a) The oscillations in $E_c$ increase as $T_D$ decreases. However, magnitude of the oscillations decreases as $T_D$ decreases, hence, there are no visible oscillations at smaller $T_D$. (b) Multidomain formations shift the conduction band downwards, hence, degrades the subthreshold performance of the device. Model is validated with the numerical COMSOL simulations [18].

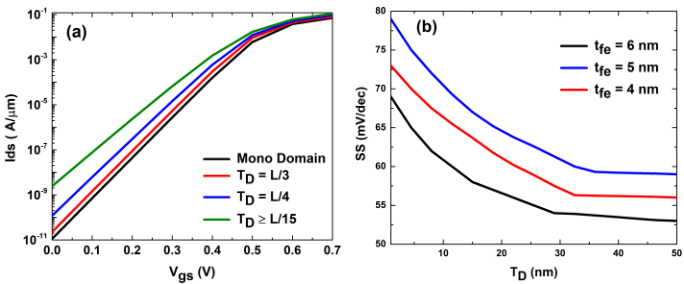

**Fig. 4:** Leakage current increases with domains concertation in the FE region (due to downward shift in the conduction band (see Fig. 3 (b)). (b) SS also degrades with the formation of multidomain, SS saturates to its minimum value, as period approaches to L.

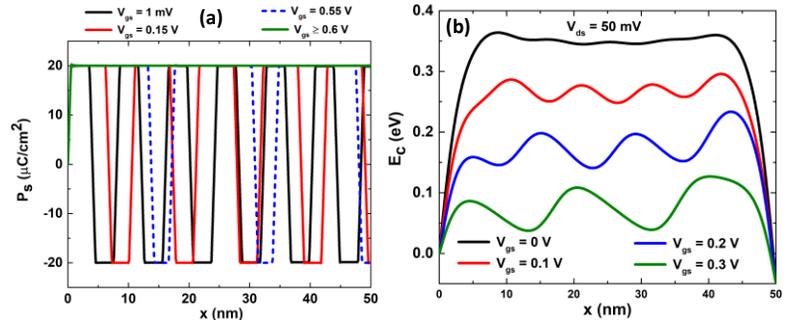

**Fig. 5:** (a) Domain switching with the gate voltage. As $V_{gs}$ increases, downward-facing domain starts to align with the applied E field, to minimize the potential energy of the system. (b) Due to the switching of domains with increase in $V_{gs}$ the period of upward domain increases, hence, rate of oscillations decreases.

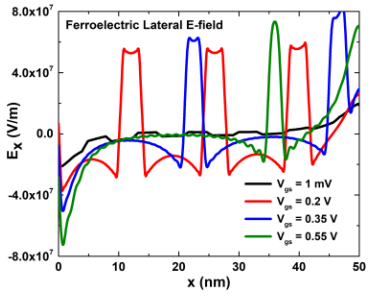 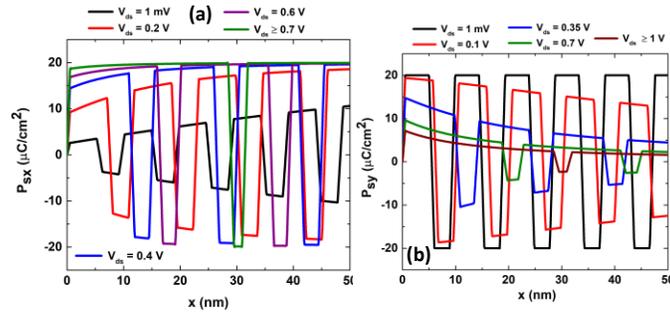 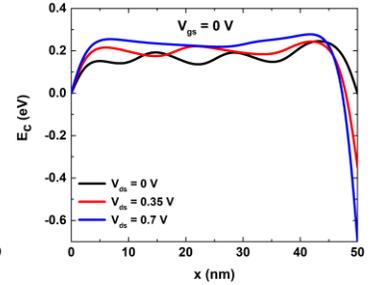

**Fig. 6:** Ferroelectric lateral direction E field exhibits the peaks with the $V_{gs}$. As $V_{gs}$ increases the number of peaks reduces due to lesser number of domains in the FE region, at the higher $V_{gs}$ (see Fig. 5(a)).

**Fig. 7:** Lateral direction polarization component with the drain voltage. The angle of $z^{th}$ domain ($\theta_z$) increases towards the drain side (see **Fig. 1**). As $V_{ds}$ increases, domains change their alignment angle to minimize the potential energy. At large $V_{ds}$ (>0.7 V) all domains align to the lateral direction. (b) The y direction component of the net polarization vector decreases, as $V_{ds}$ increases, due to increase in $\theta_z$.

**Fig. 8:** Due to the increase in $\theta_z$ with drain voltage, domains start to align in the x direction, at very large drain bias all domains align to x-axis device acts as a monodomain MFIS-NCFET.

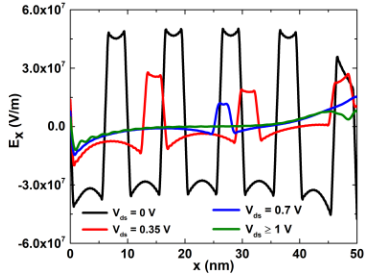 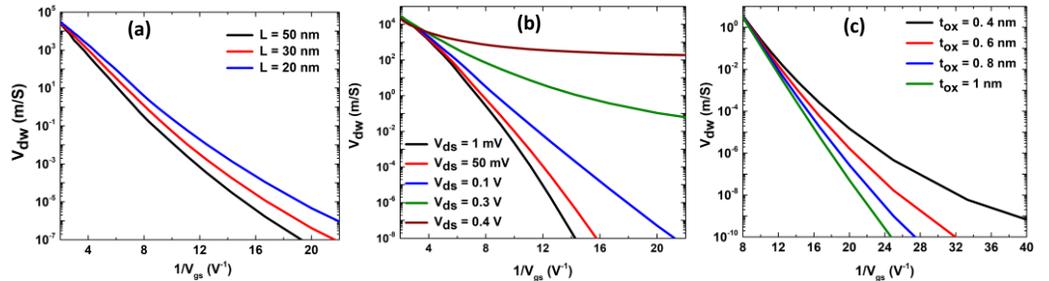

**Fig. 9:** Impact of drain bias on the FE electric field. Like Fig. 7 observations, as drain bias increases domains start to align with the net E field, and hence, oscillations in the E field decreases. At larger drain bias, all domains align in the lateral direction and form a monodomain structure.

**Fig. 10:** (a) As channel length increases, domain wall velocity decreases due to reduction in the net 2-D electric field. (b) As the drain voltages rises the lateral direction electric field increases, which helps in the rapid switching of the domain. Hence, domain wall velocity increases drain bias. (c) Larger insulator thickness decreases, the domain wall velocity due to reduction in the y-directional electric field.

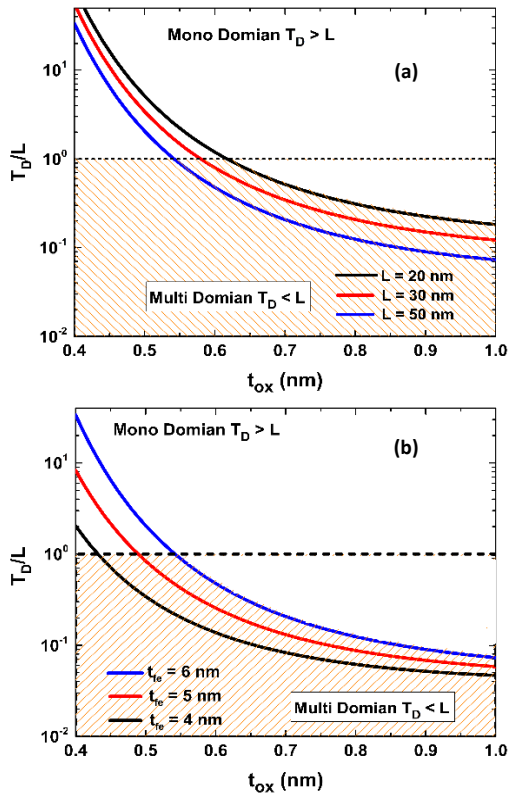

**Fig. 11:** An abrupt transition from monodomain to multidomain state is observed with increase in insulator thickness. (a) The multidomain state is more probable in the long channel device compared to the short channel device. (b) Larger $t_{fe}$ devices are more favorable to the monodomain state.

## REFERENCES

[1] M. E. Lines and A. M. Glass, Principles and Applications of Ferroelectrics and Related Materials (Oxford, 1977), chapter 4.
[2] S. Salahuddin, and S. Datta, Nano Lett. 8, 405 (2008).
[3] H. Ota *et al.*, "Fully Coupled 3-D Device Simulation of Negative Capacitance FinFETs for Sub 10 nm Integration," IEDM, Tech. Dig., 2016, pp. 12.4.1-12.4.4. [4] A. I. Khan *et al.*, "Negative Capacitance in Short-Channel FinFETs Externally Connected to an Epitaxial Ferroelectric Capacitor," EDL, vol. 37, no. 1, pp. 111-114, Jan. 2016.
[5] Cao, W *et al.*, Is negative capacitance FET a steepslope logic switch?. Nat Commun 11, 196 (2020). [6] V. P.-H. Hu *et al.*,"Negative capacitance enables FinFET and FDSOI scaling to 2 nm node,"IEDM,Tech. Dig., 2017, pp. 23.1.1-23.1.4.
[7] G. Pahwa, *et al.*,""Numerical investigation of shortchannel effects in negative capacitance MFIS and MFMIS transistors: Subthreshold behavior", TED, vol. 65, no. 11, pp. 5130-5136, Nov. 2018.
[8] Liu, S *et al*, ''Intrinsic ferroelectric switching from first principles'', Nature 534, 360–363 (2016).
[9] Saha, A.K *et al*., "Multi-Domain Negative Capacitance Effects in Metal-Ferroelectric-Insulator-Semiconductor/Metal Stacks:A Phase-field Simulation Based Study". *Sci Rep* **10,** 10207 (2020).
[10] S. Kasamatsu *et al*, "Emergence of negative capacitance in multidomain ferroelectric—Paraelectric nanocapacitors at finite bias," Adv. Mater., vol. 28, no. 2, pp. 335–340, 2016.
[11] M. Hoffmann *et al*, "Ferroelectric negative capacitance domain dynamics," J. Appl. Phys., vol. 123, no. 18, Apr. 2018, Art. no. 184101.
[12] A. Cano and D. Jim´enez, "Multidomain ferroelectricity as a limiting factor for voltage amplification in ferroelectric field-effect transistors,"
Appl. Phys. Lett., vol. 97, no. 13, p. 133509, 2010.
[13] Shin, Y *et al,* "Nucleation and growth mechanism of ferroelectric domain-wall motion", *Nature* **449,** 881–884 (2007).
[14] Park, H. W *et al,* ''Modeling of negative capacitance in ferroelectric thin flms'', Adv. Mater. 31, 1805266, (2019).
[15] N. Pandey *et al,* "Modeling of short-channel effects in DG MOSFETs: Green's function method versus scale length model," TED, vol. 65, no. 8, pp. 3112–3119, Aug. 2018.
[16] Bratkovsky, A. M.; Levanyuk, A. P. Phys. ReV. Lett. 2000, 84, 3177.
[17] L. D. Landau and E. M. Lifshitz, *Electrodynamics of Continuous Media* (Elsevier, New York, 1985), Secs. 5, 10,and 19.
[18] COMSOL Multiphysics R v. 5.4, Feb. 2018, [online] Available: http://www.comsol.com.